\documentclass[aps,superscriptaddress,twocolumn,pre,longbibliography,floatfix]{revtex4-1}

\usepackage{amsmath,amsthm,amsfonts,amssymb,bm,graphicx,color,mathpazo,times, braket}
\usepackage[colorlinks={true}, citecolor={blue}, filecolor={blue}, linkcolor={blue}, urlcolor={blue}]{hyperref}
\usepackage[caption=false]{subfig}
\usepackage{graphicx}
\usepackage{graphicx}
\usepackage{soul}
\usepackage{amsmath}

\newcommand{\sx}[0]{\hat{\sigma}_\mathrm{x}}

\newcommand{\sz}[0]{\hat{\sigma}_\mathrm{z}}

\allowdisplaybreaks

\begin{document}
\title{Quantum metrology of a structured reservoir}
	\author{Youssef Aiache}
\email{youssefaiache0@gmail.com}
\affiliation{Laboratory of R\&D in Engineering Sciences, Faculty of Sciences and Techniques Al-Hoceima, Abdelmalek Essaadi University, BP 34. Ajdir 32003, Tetouan, Morocco}

\author{Asghar Ullah}
\email{aullah21@ku.edu.tr}
\affiliation{Department of Physics, Ko\c{c} University, 34450 Sar\i yer, Istanbul, T\"urkiye}

\author{\"Ozg\"ur E. M\"ustecapl\i o\u glu}	
	\affiliation{Department of Physics, Ko\c{c} University, 34450 Sar\i yer, Istanbul, T\"urkiye}
	\affiliation{T\"UBİTAK Research Institute for Fundamental Sciences (TBAE), 41470 Gebze, T\"urkiye}
    \author{Abderrahim El Allati}
\affiliation{Laboratory of R\&D in Engineering Sciences, Faculty of Sciences and Techniques Al-Hoceima, Abdelmalek Essaadi University, BP 34. Ajdir 32003, Tetouan, Morocco}
 \date{\today} 

\begin{abstract}

Accurately characterizing the properties of structured reservoirs is a key challenge in quantum systems and is of great importance for advances in quantum metrology and sensing. In this work, we employ a two-level system (qubit) as a probe, which is coupled to a structured reservoir consisting of an ancilla qubit and a Markovian environment modeled as a thermal bath. By exploiting non-Markovian dynamics, we systematically investigate the effectiveness of different interaction types between the probe and ancilla for estimating critical parameters, including temperature, ancilla frequency, and system-bath coupling strength. We quantify the precision of parameter estimation using quantum Fisher information (QFI) and analyze the system dynamics in both transient and steady-state regimes. Our findings demonstrate that non-Markovianity substantially enhances parameter estimation in the transient regime, with specific interactions facilitating sustained information backflow and yielding higher QFI values. However, the performance of these interactions is contingent on the parameter under estimation and the operational regime. For instance, certain interactions become prominent in the transient regime but exhibit diminished utility in the steady state, whereas others maintain their effectiveness even at equilibrium. These results stress the importance of judiciously selecting interactions adapted to specific estimation objectives and operational regimes. 

\end{abstract}	
\maketitle
\section{Introduction}
Quantum metrology is one of the available practical applications of quantum technologies~\cite{RevModPhys.89.035002,BELENCHIA20221,Kurizki} and it further allows for ultra-high precision tests of fundamental physical theories~\cite{PhysRevA.108.010101,Tóth_2014}. By exploiting quantum phenomena, it enhances measurement resolution and enables unprecedented precision in parameter estimation~\cite{Giovannetti2011,RevModPhys.90.035005}. Leveraging unique quantum features such as entanglement and coherence, quantum metrology surpasses classical limits, driving advancements across diverse fields, including atomic clocks~\cite{PhysRevLett.82.2207,RevModPhys.87.637,RevModPhys.83.331,PhysRevA.54.R4649}, gravitational wave detection~\cite{Schnabel2010,Abadie2011,RevModPhys.86.121}, quantum imaging~\cite{RevModPhys.71.1539,PhysRevA.94.032301,PhysRevX.6.031033}, and even quantum biology~\cite{PhysRevX.4.011017,TAYLOR20161,Crespi}.

Recent advancements in quantum metrology have been driven by the development of advanced quantum sensors and precision measurement techniques, pushing the limits of sensitivity and accuracy~\cite{Giovannetti2011,Tóth_2014,montenegro2024}. The use of spin-squeezed states has significantly enhanced the stability of atomic clocks~\cite{PhysRevLett.125.210503,Schulte2020,PhysRevLett.117.143004}. While in gravitational wave detection, quantum-enhanced interferometry has achieved unparalleled sensitivity by employing squeezed light techniques to surpass the standard quantum limit~\cite{PhysRevLett.123.231107, Goda2008}. Beyond these applications, quantum metrology plays a pivotal role in other subfields, such as quantum magnetometry~\cite{Jonathan,PhysRevLett.104.133601}, quantum thermometry~\cite{Mehboudi_2019,DePasquale2018}, and gravimetry, among others. In particular, temperature sensing has gained significant attention in recent years~\cite{Mehboudi_2019,Giovannetti2011}. From an application perspective, quantum devices such as cold-atom and ion-trap systems often operate at extremely low temperatures~\cite{PhysRevLett.125.080402,PhysRevX.10.011018,PhysRevLett.122.030403,PhysRevA.88.063609,Olf2015}. Consequently, precise temperature measurements are essential not only for maintaining optimal performance but also for harnessing quantum effects to enhance precision and sensitivity in sensing~\cite{PhysRevA.98.050101,PhysRevA.101.032112,PhysRevResearch.2.033394,PhysRevA.99.062114,PhysRevA.96.062103,PhysRevLett.114.220405,PhysRevA.91.012331,PhysRevA.84.032105,PhysRevA.82.011611,PhysRevLett.125.080402,Potts2019fundamentallimits,Mukherjee2019,Campbell_2017,PhysRevLett.128.040502,Purdy,PhysRevResearch.5.043184}.

In general, a quantum system can serve as a probe to measure the physical parameters of an environment in both thermal equilibrium and non-equilibrium settings~\cite{PhysRevResearch.1.033021,PhysRevA.96.012316,PhysRevA.99.062114,PhysRevA.108.062421}. Quantum probes, such as spins, atoms, or photons, interact with the environment in a way that encodes information about its physical parameters. By carefully designing these interactions, quantum probes can extract maximum information about the environment, even in the presence of noise and decoherence~\cite{Zhang2022,PhysRevA.110.062406,He2023}. While fundamental limits, such as the Heisenberg uncertainty principle, constrain the precision of measurements~\cite{Giovannetti2011,RevModPhys.90.035005,Tóth_2014,PhysRevLett.105.180402}, quantum strategies allow us to approach these limits more closely than classical methods.
Structured reservoirs, such as thermal baths with complex interactions and environmental couplings, provide a rich platform for parameter estimation in quantum systems~\cite{PhysRevA.97.012125,e21050486}. These systems often exhibit interdependent parameters, such as coupling strengths, decay rates, and temperature gradients, whose precise determination is essential for understanding and controlling quantum dynamics~\cite{Nokkala2016}.

Quantum probes, such as qubits, offer a practical means to estimate the unknown parameters by interacting with the reservoir and encoding its properties into measurable quantities. In this work, we employ a two-level system as a probe coupled to a structured reservoir, which consists of an ancilla qubit and a Markovian environment (a thermal bath). Typically, structured reservoirs are characterized by spectral density functions composed of Lorentzians, defined by a central frequency, bandwidth, and coupling strength. In our work, we model such structured environments using a Markovian bath coupled to an ancilla qubit, where the qubit's frequency acts as a color filter, effectively capturing the central frequency of the Lorentzian, while the bath temperature and interaction strength parametrize the environment, enabling the probe to capture its essential features. This setup induces non-Markovian dynamics, providing a unique opportunity to enhance parameter estimation beyond traditional Markovian models~\cite{PhysRevLett.109.233601, Berrada:17}. By leveraging these memory effects, we explore how different types of interactions between the probe and the reservoir can optimize the extraction of information.
We quantify the precision of parameter estimation using quantum Fisher information (QFI)~\cite{paris2009}, which sets a fundamental bound on the precision of parameter estimation in quantum systems~\cite{Tóth_2014,Liu_2020}. Our analysis employs the Lindblad master equation to capture both the dissipative dynamics and the system-bath interaction, enabling a comprehensive investigation of parameter estimation in both transient and steady-state regimes. We focus on estimating key structured reservoir parameters, including the temperature of the bath, ancilla frequency, and system-bath coupling strength.

Our results reveal that non-Markovianity significantly enhances parameter estimation in the transient regime, with certain interactions enabling sustained information backflow and results in higher values of QFI. However, the effectiveness of these interactions varies depending on the parameter being estimated and the regime of operation. For instance, some interactions are effective in the transient regime but lose their effectiveness in the steady state, while others retain their utility even at equilibrium. 
We also investigate how coherence generated in the probe affects parameter estimation. Our results show that coherence, when preserved by specific interactions, improves estimation accuracy by encoding additional information beyond the population dynamics. This effect is particularly significant for interactions that allow coherence to actively contribute to the dynamics of the system, thereby improving the accuracy of parameter estimation.

The rest of the paper is organized as follows: In Sec.~\ref{brief_review}, we briefly review key concepts, including structured reservoirs and the QFI. In Sec.~\ref{model}, we introduce the system model and describe the dynamics of the probe qubit coupled to the structured reservoir, and then discuss non-Markovianity. Section.~\ref{Results} presents our results on temperature estimation, followed by the estimation of the ancilla qubit's frequency and its coupling strength with the reservoir. We then analyze the steady-state results for parameter estimation in the structured reservoir. Concluding remarks are given in Sec.~\ref{conclusion}. The optimal measurements for extracting information about each parameter are discussed in Appendix~\ref{Optimal_measurements}, while the role of coherence is explored in Appendix~\ref{coh}.
\section{Theoretical background}
\label{brief_review}
\subsection{Structured reservoirs}
In open quantum systems, the environment cannot be accurately described by a memoryless, white-noise (Markovian) bath. Instead, it often exhibits a structured spectral response, leading to non-Markovian dynamics with temporal correlations and information backflow. Such environments are termed structured reservoirs~\cite{PhysRevA.97.032133,PhysRevA.102.012217}. Modeling non-Markovian (colored) environments is essential for understanding open quantum systems beyond the Markovian approximation. A physically transparent and elegant approach is to couple the system to an intermediate quantum mode (e.g., a two-level atom or harmonic oscillator), which itself interacts with a Markovian heat bath. This intermediate system acts as a spectral filter, transforming the flat spectral density of the bath into a structured (e.g., Lorentzian) spectrum experienced by the system.

We consider a system qubit $S$, and an auxiliary two-level atom $A$ with Hamiltonians
  \begin{equation}
\begin{aligned}
      \hat{H}_S = \frac{\omega_S}{2} \hat{\sigma}_z^S, \quad \hat{H}_A = \frac{\omega_A}{2} \hat{\sigma}_z^A.
\end{aligned}
\end{equation}
A Markovian bosonic bath $B$ is given by the Hamiltonian
  \begin{equation}
  \hat{H}_B = \sum_k \omega_k b_k^\dagger b_k.
  \end{equation}
The interactions between the system and the auxiliary, and between the auxiliary and the bath, are given by
\begin{align}
\hat{H}_{SA} &= g(\hat{\sigma}_+^S \hat{\sigma}_-^A + \hat{\sigma}_-^S \hat{\sigma}_+^A) \\
\hat{H}_{AB} &= \sum_k (h_k \hat{\sigma}_+^A \hat{b}_k + h_k^* \hat{\sigma}_-^A \hat{b}_k^\dagger).
\end{align}
Here, $g$ is the flip-flop coupling between the system and the atom, and $h_k$ are the atom-bath coupling constants. We assume that the bath has a flat spectral density, such as
\begin{equation}
J_B(\omega) = \sum_k |h_k|^2 \delta(\omega - \omega_k) \approx \gamma_A,
\end{equation}
where $\gamma_A$ is the decay rate of the atom due to the bath. Tracing out the bath, the atom’s susceptibility becomes
\begin{equation}
\chi_A(\omega) = \frac{1}{\omega_A - \omega - i \gamma_A/2},
\end{equation}
which describes a Lorentzian frequency response centered at $\omega_A$ with linewidth $\gamma_A$.
The effective spectral density seen by the system qubit $S$ is then
\begin{equation}
J_{\text{eff}}(\omega) \propto |g|^2 \cdot \frac{\gamma_A/2}{(\omega_A - \omega)^2 + (\gamma_A/2)^2},
\end{equation}
which is a Lorentzian spectrum, characteristic of a colored, structured environment. The reduced dynamics of the system qubit $S$ can be described by a generalized master equation
\begin{equation}
\frac{d}{dt} \rho_S(t) = -i[\hat{H}_S, \rho_S(t)] + \int_0^t K(t - \tau) \rho_S(\tau) d\tau,
\end{equation}
where $K(t)$ is the memory kernel, which for a Lorentzian spectral density approximately takes the form
\begin{equation}
K(t) \propto e^{-\gamma_A t/2} e^{-i\omega_A t},
\end{equation}
reflecting the finite correlation time $\tau_c \sim 2/\gamma_A$ of the effective bath. This approach is closely related to the pseudomode formalism~\cite{PhysRevA.55.2290}, reaction coordinate mapping~\cite{PhysRevA.90.032114, Strasberg_2016}, and spectral filtering methods~\cite{Gelbwaser-Klimovsky2015, Kofman01021994}, which represent structured environments via discrete auxiliary modes. 

By coupling a system qubit to an intermediate two-level atom connected to a Markovian bath, one can effectively simulate non-Markovian dynamics with a structured, Lorentzian spectral density. This approach provides a powerful, minimal model for exploring non-Markovian open system dynamics in
quantum thermodynamics, metrology, and information processing.

\subsection{Fundamentals of parameter estimation}
Consider a \(d\)-dimensional parametrized state \(\rho_{\boldsymbol{\theta}}\), where \(\boldsymbol{\theta} = (\theta_1, \theta_2, \dots, \theta_k)\) represents \(k\) unknown parameters to be estimated. In the general multiparameter estimation scenario, the performance of an estimator \(\boldsymbol{\hat{\theta}}\) is characterized by the Cram\'er--Rao bound (CRB), which we briefly outline here for completeness. However, in our current scheme, we assume that the parameters $\omega_A$, $\gamma$, and $T$ can be estimated independently. This is justified by the assumption that when estimating a given parameter (e.g., $\omega_A$), the others (e.g., $\gamma$ and $T$) are known exactly from prior knowledge. This allows us to treat the problem as a sequence of single-parameter estimation tasks, rather than a simultaneous multiparameter problem. Therefore, although we summarize the general multiparameter formalism below, we focus only on single-parameter estimation in our results.
\begin{figure}[t!]
    \begin{center}
	\includegraphics[scale=.43]{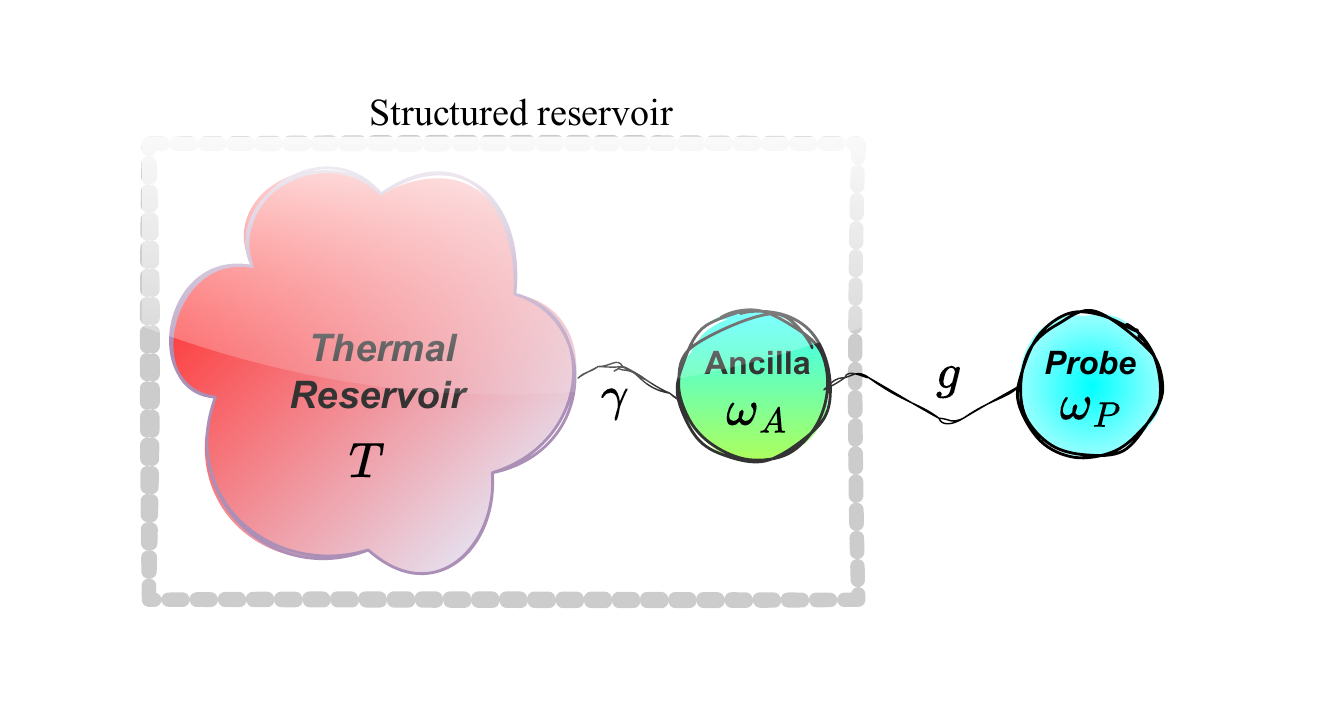}
	\caption{Our scheme consists of two qubits: an ancilla qubit with transition frequency $\omega_A$, directly coupled to a thermal bath at temperature $T$. Moreover, we consider a probe qubit with transition frequency $\omega_P$, which has no direct access to the bath. The ancilla and the bath together form a structured reservoir, which acts as an effective environment for the probe qubit. The probe system interacts with the ancilla via an interaction Hamiltonian $\hat{H}_{AP}$, allowing it to infer different parameters of the structured reservoir. The probe qubit serves as the measurement system, enabling parameter estimation through energy measurements performed on it.}
	\label{fig:1}
    \end{center}
\end{figure}
The CRB for the multiparameter case (for single \(m = 1\) measurement) reads~\cite{paris2009,Sidhu,Liu_2020}
\begin{equation}\label{CRB}
    \text{Cov}[\boldsymbol{\hat{\theta}}]\ge\mathcal{F}_C(\boldsymbol{\theta})^{-1},
\end{equation}
where $\text{Cov}[\boldsymbol{\hat{\theta}}]$ is the covariance matrix with elements
\begin{equation}
    [\text{Cov}(\boldsymbol{\hat{\theta}})]_{ab}=\langle\theta_a\theta_b\rangle-\langle\theta_a\rangle\langle\theta_b\rangle,
\end{equation}
and $\mathcal{F}_C(\boldsymbol{\theta})$ is the classical FI matrix whose diagonal elements satisfy the bound $\text{Var}[\hat{\theta}_a]\ge[\mathcal{F}_C(\hat{\theta})]_{aa}$. Similarly to the single-parameter scenario, it is possible to obtain a tighter bound of Eq.~\eqref{CRB}, which leads to the multiparameter quantum CRB~\cite{paris2009,Sidhu,Liu_2020}
\begin{equation}
    \text{Cov}[\hat{\theta}]\ge\mathcal{F}_C(\boldsymbol{\theta})^{-1}\ge\mathcal{F}_Q(\boldsymbol{\theta})^{-1},
\end{equation}
where the matrix elements of the QFI matrix $\mathcal{F}_Q(\boldsymbol{\theta})$ are given by
\begin{equation}\label{qfi}
    [\mathcal{F}_Q(\boldsymbol{\theta})]_{ab}=\frac{1}{2}\text{Tr}[\rho_{\boldsymbol{\theta}}\{L_a,L_b\}],
\end{equation}
where $\{.,.\}$ denotes the anti-commutator and $L_a(L_b)$ is the symmetric logarithmic derivative (SLD) for a parameter $\theta_a(\theta_b)$ which is determined by the equation
\begin{equation}
      \frac{\partial\rho_{\theta_a}}{\partial\theta_a} = \frac{L_{\theta_a}\rho_{\theta_a} + \rho_{\theta_a} L_{\theta_a}}{2}.
\end{equation}
The SLD is a Hermitian operator whose expected value is $\text{Tr}(\rho L_a)=0$. Based on Eq. \eqref{qfi}, the diagonal element of the QFI matrix is $\mathcal{F}_{aa}=\text{Tr}(\rho L_a^2)$ which is exactly the QFI for a parameter $\theta_a$ whereas the off-diagonal entry is given by $\mathcal{F}_{ab}=\text{Tr}(L_b\partial_a\rho)$~\cite{Liu_2020}. The notation $\partial_{\theta}$ is used as an abbreviation of $\partial/\partial\theta$.

Given this assumption of parameter separability, we now focus on single-parameter estimation and apply the corresponding expression for the QFI. Recall that in the Bloch sphere representation, any qubit state can be written as

\begin{equation}
    \rho = \frac{1}{2} \left( \mathbb{I} + \mathbf{r} \cdot \boldsymbol{\sigma} \right),
\end{equation}
where $\mathbf{r}=(r_x,r_y,r_z)^T$ is the real Bloch vector, $\boldsymbol{\sigma}=(\sigma_x,\sigma_y,\sigma_z)$ represents the spin-1/2 Pauli matrices, and $\mathbb{I}$ is $2\times2$ the identity matrix. For a mixed quantum state, one can calculate the QFI using the following expression~\cite{PhysRevA.87.022337}
\begin{equation}
\mathcal{F}_\theta=|\partial_{\theta}\mathbf{r}|^2+\frac{(\mathbf{r}\cdot\partial_\theta\mathbf{r})^2}{1-|\mathbf{r}|^2},
\end{equation}

\section{model and system dynamics}
\label{model}
We detail our model of the engineered structured reservoir for quantum metrology, which induces a
non-Markovian dynamics on the probe qubit. The reservoir comprises two parts, a Markovian thermal reservoir of temperature $T$ and a two-level system (henceforth referred to as ancilla qubit), as depicted in Fig.~\ref{fig:1}. The probe qubit accesses
the reservoir indirectly through the ancilla, which mediates the interaction.

The Hamiltonian of the probe-ancilla system is given as 
\begin{eqnarray}\label{H_S}
		\hat{H}_S&=&\dfrac{\hbar}{2}\omega_P\sz^P+\dfrac{\hbar}{2}\omega_A\sz^A+\hat{H}^{(i)}_{AP},
\end{eqnarray}
where $\omega_P$ and $\omega_A$ denote the frequencies of the probe and ancilla qubits, respectively, and \( \sigma_z^{P(A)} \) represents the \( z \)-component of the Pauli matrix for the probe (ancilla) qubit. The third term in Eq.~(\ref{H_S}) corresponds to the interaction Hamiltonian between the probe and ancilla qubits. Throughout this work, we individually investigate four families of interactions $\hat{H}_{AP}$ terms between the ancilla and probe, as detailed below~\cite{Cahit,Kolář_2024}
\begin{eqnarray}\label{H_AP}
		\hat{H}^{(i)}_{AP}=
		\begin{cases}
			\hat{H}^{(XX)}_{AP}&=g~\sx^P\sx^A, \\
			\hat{H}^{(XX+ZX)}_{AP}&=g~\big(\sx^P\sx^A+\sz^P\sx^A\big), \\
			\hat{H}^{(ZX)}_{AP}&=g~\sz^P\sx^A, \\
			\hat{H}^{(XZ)}_{AP}&=g~\sx^P\sz^A.
		\end{cases}
\end{eqnarray}
From a practical standpoint, the interaction types are motivated by their appearance in experimentally realizable quantum systems, particularly in superconducting circuit QED architectures~\cite{PhysRevA.69.062320}. In these setups, qubits are often coupled to resonators or shared circuit elements, resulting in effective spin-spin interactions once the mediating degrees of freedom are integrated out~\cite{RevModPhys.93.025005}. The precise form of the coupling depends on the circuit design and the symmetry of the qubit-resonator interaction~\cite{PhysRevLett.105.237001, PhysRevA.69.062320, PhysRevLett.105.060503}. In addition, certain circuit QED architectures allow for switchable interactions between qubits, where all three coupling orientations ($\hat{\sigma}^x$, $\hat{\sigma}^z$, $\hat{\sigma}^z$ ) can be engineered. As shown in Ref.~\cite{PhysRevLett.105.023601}, the interaction orientation in the qubit basis can be dynamically tuned using flux-controlled designs. Together, these features render such interactions not only experimentally relevant but also widely realizable across different quantum platforms.
    \begin{figure*}[t!]
		\centering
		\includegraphics[width=1\textwidth]{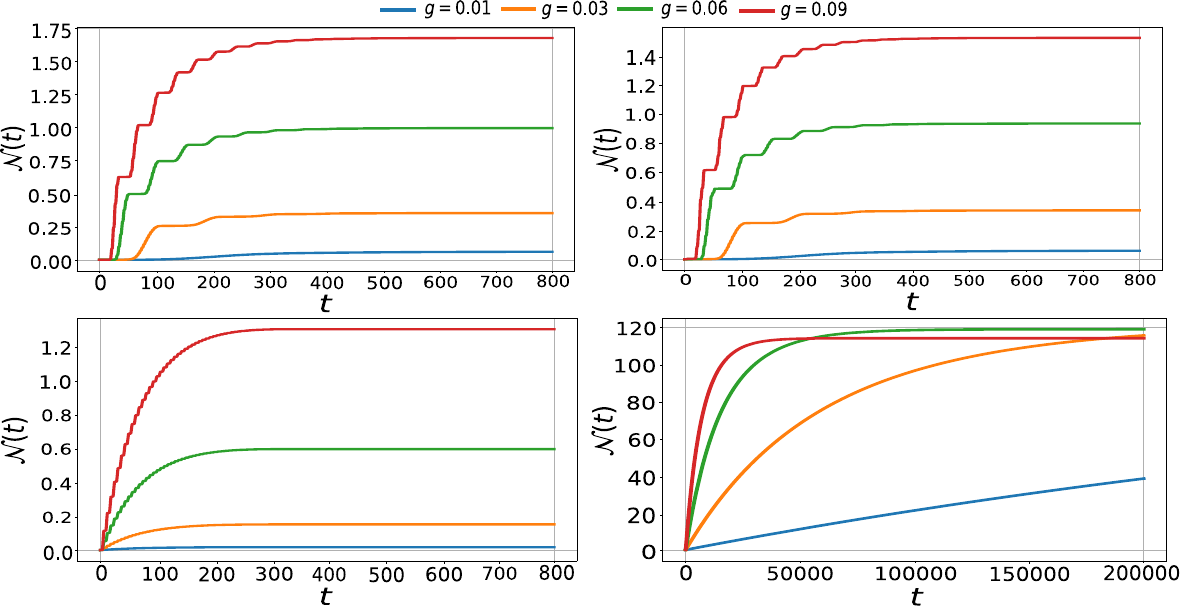}
        \put(-300,210){${(\textbf{a})}$}~
	\put(-50,210){${(\textbf{b})}$}~
	\put(-300,100){${(\textbf{c})}$}~
	\put(-50,100){${(\textbf{d})}$}~
		\caption{ Non-Markovianity measure $\mathcal{N}(t)$ as a function of time $t$ for different models as described in Eq. \eqref{H_AP} . The results are obtained for different types of interaction Hamiltonians such as: (\textbf{a}) $ H^{(XX)}_{AP} $, (\textbf{b}) $ H^{(XX+ZX)}_{AP} $, (\textbf{c}) $ H^{(ZX)}_{AP} $, and (\textbf{d}) $ H^{(XZ)}_{AP} $. The blue, orange, green, and red lines represent coupling strengths $ g = 0.01,~ 0.03,~ 0.06, $ and $ 0.09 $, respectively. The rest of the parameters are set to $ \omega_A = 0.99 $, $ \omega_P = 1 $, $ T = 0.3 $, and $ \gamma = 0.05 $.}
	\label{fig:2}
    \end{figure*}

The ancilla interacts with the reservoir according to the following linear dissipative interaction Hamiltonian~\cite{Breuer,rivas2012open}
	\begin{eqnarray}
	   \hat{H}_{AR}=\sx^A\otimes\sum_{k}g_k( \hat{b}_k^{\dagger}+ \hat{b}_k).
	\end{eqnarray}
For concreteness, we model the ancilla reservoir as a composite system in which only the ancilla directly interacts with the thermal bath with coupling strength \(\gamma\), while the probe is coupled solely to the ancilla. Under conditions that validate the Born-Markov approximation, the evolution of the density matrix of the system \(\rho_S(t)\) follows the Lindblad master equation~\cite{Breuer}

\begin{align}\label{eq:ME}
    \frac{d\rho_S(t)}{dt}= \mathcal{L}_T\rho_S(t),
\end{align}
where the Liouvillian super-operator is given by  
\begin{align}\label{eq:LT}
    \mathcal{L}_T = -i[\hat{H}_S,\rho_S(t)]
    + \gamma_-\mathcal{D}[\hat{\sigma}_-^A]\rho_S(t)
    + \gamma_+\mathcal{D}[\hat{\sigma}_+^A]\rho_S(t).
\end{align}
Here, the dissipator is defined as  
\begin{align}
    \mathcal{D}[\hat{A}]\rho_S(t) = \hat{A}\rho_S(t)\hat{A}^\dagger - \frac{1}{2} \{\hat{A}^\dagger \hat{A},\rho_S(t)\},
\end{align}
where \(\{\cdot,\cdot\}\) denotes the anti-commutator. The decay rates appearing in Eq.~(\ref{eq:LT}), are expressed as  
\begin{align}
    \gamma_- = (n_{th} + 1)\gamma \quad \text{and} \quad \gamma_+ = n_{th} \gamma,
\end{align}
where \(n_{th}\) is the average number of resonant thermal excitations, given by the Bose-Einstein distribution
\begin{align}\label{eq:bose}
    n_{th} = \frac{1}{e^{\omega_A/T} - 1}.
\end{align}
Thus, the temperature dependence of the system is fully incorporated through the ancilla's interaction with the thermal bath. 

Our primary focus is on the state of the probe, which remains isolated from direct interaction with the thermal reservoir. We trace out the degrees of freedom of the ancilla to obtain the reduced state of the probe. This enables us to extract parameter information solely from the probe's measurements. The reduced density operator of the probe is given by  
\begin{eqnarray}\label{Probe_state}
    \rho_P(t) = \text{Tr}_A\big\{ \rho_S(t) \big\},
\end{eqnarray}
where \(\rho_P(t)\) represents the state of the quantum probe, and \(\text{Tr}_A\) denotes the partial trace over the ancilla's degrees of freedom.

\subsection{Non-Markovianity}
In quantum metrology, the precision of parameter estimation depends critically on how the probe system exchanges information with its environment. In conventional Markovian dynamics, this exchange is unidirectional information continuously dissipates into the environment. However, when memory effects are present, i.e., in non-Markovian dynamics, information can temporarily flow back to the probe, potentially enhancing the distinguishability of quantum states and thus improving metrological performance. In our model, the presence of an ancilla that mediates the interaction between the probe and the bath introduces structured dynamics that can lead to such non-Markovian effects ~\cite{PhysRevA.102.012217,PhysRevE.110.024132}. Therefore, quantifying the degree of non-Markovianity becomes essential for understanding the probe’s behavior and its sensitivity to different parameters. To this end, we employ a widely used witness of non-Markovianity based on the trace distance between two evolving quantum states, as introduced by Breuer \textit{et al}.\cite{PhysRevLett.103.210401}, that captures the flow of information between the probe and its environment and allows us to identify regimes where non-Markovianity may contribute positively to probe’s performance.
However, since multiple probe-ancilla interactions are considered, as described in Eq.~(\ref{H_AP}), we analyze the non-Markovianity case by case. The non-Markovianity measure, denoted by $\mathcal{N}(t)$, is given by the compact expression
	\begin{equation}\label{N}
		\mathcal{N}(t) = \max_{\rho_P^{(1)}(0), \rho_P^{(2)}(0)} \int_{\sigma > 0} \sigma(t) dt,
	\end{equation}
where $\sigma(t)$ represents the rate for which the trace distance  changes over time between two distinct quantum states, \( \rho_P^{(1)}(t) \) and \( \rho_P^{(2)}(t) \), with initial states \( \rho_P^{(1)}(0) \) and \( \rho_P^{(2)}(0) \), respectively,  defined as
\begin{equation}
			\sigma(t) = \frac{d}{dt} \mathcal{D}[\rho_P^{(1)}(t), \rho_P^{(2)}(t)],
\end{equation}
The trace distance expresses our ability to distinguish the states and is defined as
\begin{equation}
    \mathcal{D}[\rho_P^{(1)}(t), \rho_P^{(2)}(t)] =\frac{1}{2}|\rho_P^{(1)}(t)-\rho_P^{(2)}(t)|,
\end{equation}
with $|A|=\text{Tr}\sqrt{A^\dagger A}$ for a square matrix $A$. In addition, we consider orthogonal pure states for a two-level quantum system as the initial states, given by $|\pm\rangle = (|0\rangle \pm |1\rangle)/\sqrt{2}$ \cite{RevModPhys.89.015001, RevModPhys.88.021002}. These states are particularly useful for capturing the maximum information backflow from the structured reservoir to the probe. Importantly, the integral in Eq.~(\ref{N}) is carried out over time intervals. Indeed, for a positive value of \( \sigma(t) \), the backflow of information from the environment to the system appears, which is a signature of non-Markovian dynamics.

\section{Results}
\label{Results}
In this section, we first discuss the parameters considered for numerical simulations and their physical importance. We then present our results for the estimation of different parameters of the structured reservoir in the following subsections. For the QFI analysis, we assume that the probe and ancilla are initially prepared in the ground ($|0\rangle$) and excited states ($|1\rangle$), respectively.
\subsection{Physical parameters}
We work in dimensionless units and scale all system parameters by the frequency $\omega_P$ of the probe qubit. Without loss of generality, we focus on the off-resonant regime $\omega_P>\omega_A$, which is more general and allows richer dynamics, while the resonant case yields qualitatively similar results. We also set $\hbar=k_B=1$ throughout the study. For concreteness, assuming $\omega_P \sim 30$ GHz, which is a typical qubit frequency in superconducting circuit platforms, the dimensionless temperature range explored here (up to $T=1$) corresponds to physical temperatures around $230$ $mK$~\cite{Marco2021}, which is well within the operational range of current circuit QED experiments. Therefore, the fixed parameters chosen in this work are consistent with experimentally accessible regimes.
\subsection{Estimation of temperature}
 In this section, we discuss our results for the estimation of temperature $T$ of the structured reservoir for different types of interactions and compare their effectiveness, taking advantage of non-Markovianity~\cite{PhysRevApplied.17.034073}. We first discuss the dynamics of non-Markovianity for different models. 
\begin{figure*}[t!]
        \includegraphics[scale=0.8]{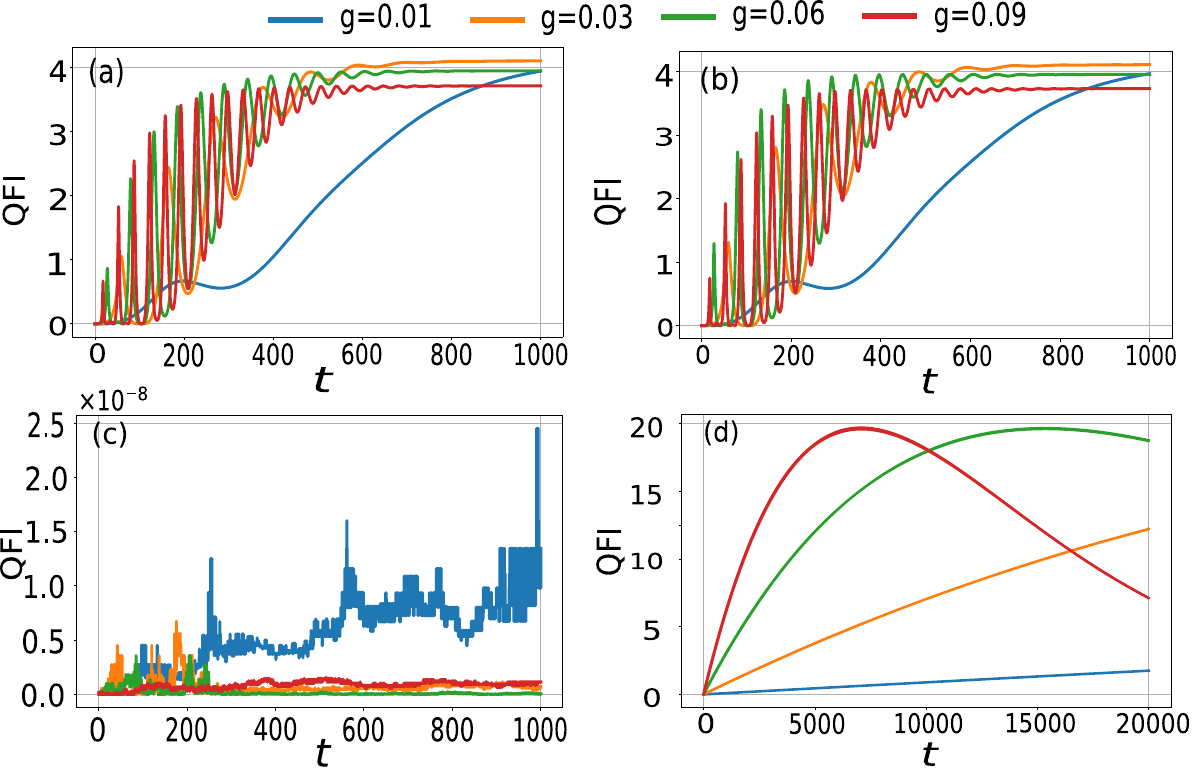}
	\caption{QFI as a function of time for the estimation of bath temperature $T$. The results are for different models as shown in eq.~\eqref{H_AP}  with (\textbf{a}) $ H^{(XX)}_{AP} $. (\textbf{b}) $ H^{(XX+ZX)}_{AP} $. (\textbf{c}) $ H^{(ZX)}_{AP} $, and (\textbf{d}) $ H^{(XZ)}_{AP} $. The blue, orange, green, and red lines correspond to $ g=0.01, 0.03, 0.06, \text{and}~ 0.09$. The rest of the  parameters are set to $\omega_A=0.99$, $\omega_P=1$, $ T =0.3$, and $ \gamma=0.05 $.}
	\label{fig:3}
\end{figure*}
 Figure~\ref{fig:2} illustrates the evolution of $\mathcal{N}(t)$ over time for different models and coupling strengths $g$. Initially, the non-Markovianity $\mathcal{N}(t)$ exhibits rapid growth, with stronger coupling strength resulting in a faster rise and a higher asymptotic value. The growth pattern varies across models: in some cases, it is step-like, signifying discrete information backflows, while in others, it is smooth. Following an initial transient phase, $\mathcal{N}(t)$ saturates, meaning no further increase occurs at large times. The saturation value is higher for stronger coupling and varies across models. Notably, the model with interaction Hamiltonian $\hat{H}_{\rm AP}^{(XZ)}$ exhibits the highest overall non-Markovianity, while other interaction Hamiltonians such as $\hat{H}_{\rm AP}^{(XX)}$$, \hat{H}_{\rm AP}^{(XX+ZX)}$, and $\hat{H}_{\rm AP}^{(ZX)}$ display similar qualitative behavior but reach lower saturation values. The differences in saturation values and the nature of the increase (step-like vs. smooth) highlight the distinct dynamical properties of each model under varying coupling strengths. Furthermore, the observed behavior of $\mathcal{N}(t)$ in these models can be leveraged to estimate different parameters of the structured reservoir, providing a practical tool for probing reservoir properties through non-Markovian dynamics using different models.

In Fig.~\ref{fig:3}, we plot QFI as a function of time for different types of interactions between the probe and ancilla qubit. We consider small detuning between the qubits, such as $\Delta\omega=0.01$, and investigate the impact of different coupling strengths on the estimation of temperature. Figure.~\ref{fig:3}(a) shows QFI for estimation of $T$ when the two qubits are coupled through interaction Hamiltonian $H^{XX}_\text{AP}$. We can see that when the coupling strength is weak such as $g=0.01$, the QFI is almost zero initially and it increases slowly with respect to time and eventually it smoothly reaches its maximum value (solid blue curve). However, if the coupling strength is strong, then the QFI abruptly increases and reaches its maximum value very fast, but this time it shows very rapid oscillations over a short interval of time. This is possible because the information backflow is in the form of step-like as shown by the non-Markovianity in Fig.~\ref{fig:2}. This discrete information backflow causes the oscillations in QFI.

Although the non-Markovianity of the interaction \( H^{ZX}_\text{AP} \) is nonzero, the corresponding QFI remains negligible, as shown in Fig.~\ref{fig:3}(b). This interaction is antisymmetric, which prevents effective temperature encoding. Additionally, it causes the probe qubit to experience a fluctuating field, leading to dephasing and the loss of useful thermal information. Consequently, temperature information is not efficiently imprinted onto the probe qubit’s state.

Therefore, if we consider \( H^{ZX}_\text{AP} \) together with the \( H^{XX}_\text{AP} \) interaction, they do not significantly impact the QFI (see Fig.~\ref{fig:3}(c)). A particularly interesting and important interaction is \( H^{XZ}_\text{AP} \), which exhibits higher values of \( \mathcal{N}(t) \) that persist for a longer duration. This indicates that information backflow remains significant over an extended period. This highlights that non-Markovianity alone does not guarantee enhanced estimation precision. Although the two interactions involve similar Pauli terms, they influence the probe's information dynamics in fundamentally different ways. The key difference lies in the directionality of the interaction. In the case of \( H^{XZ}_\text{AP} \), the probe couples via \( \hat{\sigma}^x \), enabling the generation and persistence of temperature-sensitive coherence (see red curve in Fig.~\ref{fig:coh} below), which directly contributes to enhanced QFI. Conversely, under \( ZX \) coupling, the probe interacts through \( \hat{\sigma}^z \), which commutes with the probe's free Hamiltonian. This suppresses the coherent evolution required for effective parameter encoding, resulting in a low QFI despite the presence of non-Markovian effects due to ancilla back-action.

Notably, the QFI takes a long time to reach its peak, but interestingly, it eventually decays to zero in the long-time limit, as shown in Fig.~\ref{fig:3}(d). These results demonstrate that among all considered interactions, \( H^{ZX}_\text{AP} \) performs the worst, whereas \( H^{XZ}_\text{AP} \) is the most effective for reservoir temperature estimation.

However, the \( H^{ZX}_\text{AP} \) interaction is ineffective in extracting frequency information; therefore, we do not report its results here. However, the \( H^{XZ}_\text{AP} \) interaction again performs well, yielding significantly higher QFI values, as illustrated in Fig.~\ref{fig:4}(a). Similar to the case of temperature estimation, the QFI for frequency estimation also requires a long time to reach its peak value before eventually decaying to zero.  
\begin{figure}[t!]
    \centering
    \includegraphics[scale=0.41]{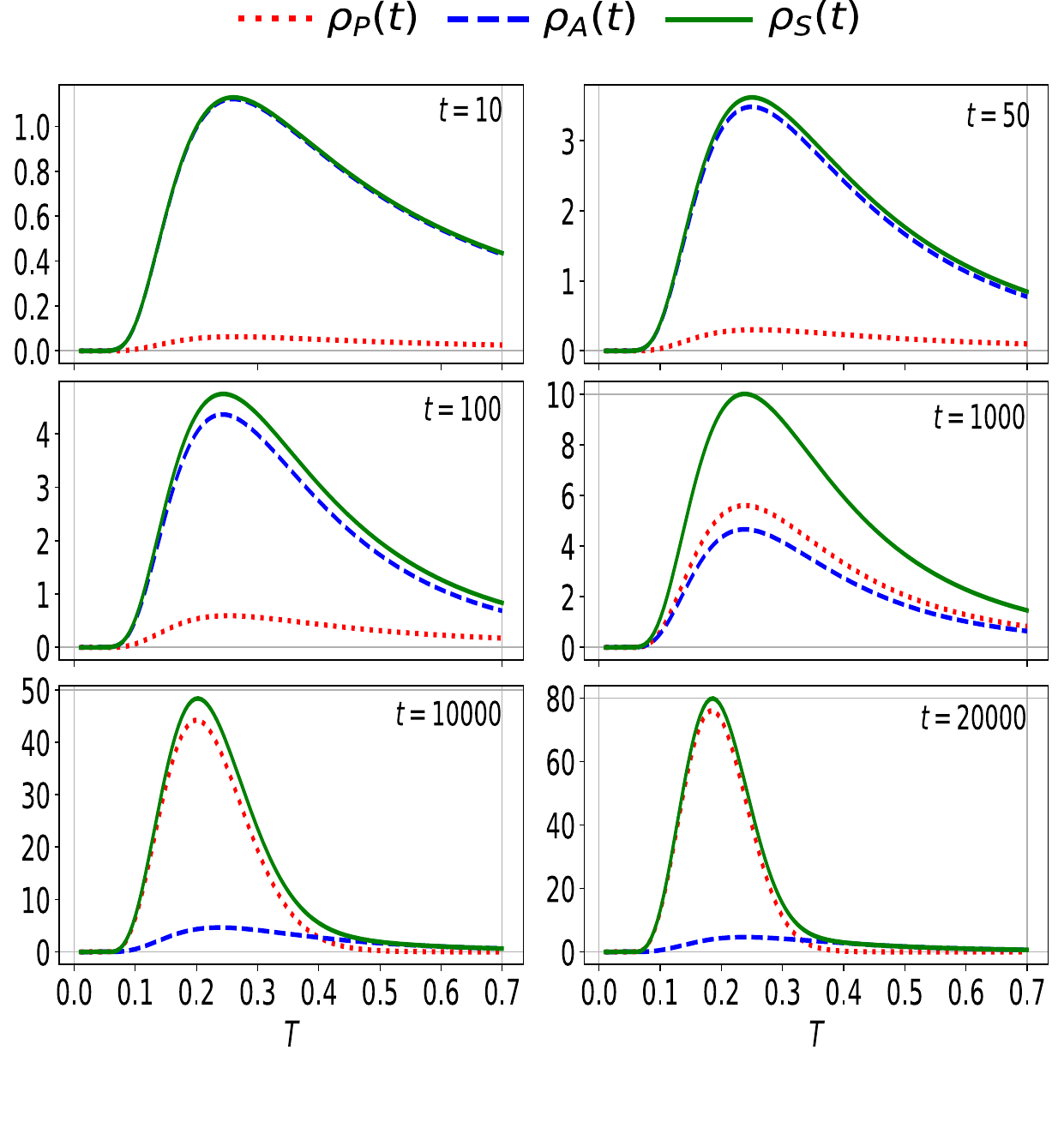}
    \caption{QFI $\mathcal{F}(t)$ associated with the full 
 state $\rho_S(t)$, the ancilla qubit state $\rho_A(t)$, and probe qubit state $\rho_P(t)$. The results are shown as a function of reservoir temperature $T$. The panels correspond to different probing times $t$ as annotated in the figure. The rest of the parameters are set to $\omega_P=1$, $\omega_A=0.99$, and $\gamma=0.05$.}
    \label{fig:prob}
\end{figure}

In Fig.~\ref{fig:prob}, we compare the QFI associated with the reduced states of either the ancilla \(\rho_A(t)\) or the probe \(\rho_P(t)\) to that of the full probe-ancilla system $\rho_S(t)$. The results are shown as a function of the reservoir temperature \(T\) for different probing times \(t\), considering the \(H^{XZ}_{AP}\) interaction between the probe and ancilla qubits in Fig.~\ref{fig:prob}. 

At short evolution times, such as \(t=10\), the thermometric performance is primarily governed by the ancilla qubit, with \(\rho_A(t) \approx \rho_S(t)\), while the probe qubit exhibits a very small QFI. As the probing time increases, temperature information gradually accumulates in the probe qubit's population through non-Markovianity. For instance, at \(t=10000\), the QFI of the probe qubit matches that of the full system, i.e., \(\mathcal{F}^P_T(t) \approx\mathcal{F}^S_T(t)\). 
  
For even longer times, such as \( t = 20000 \), the temperature information becomes predominantly stored in the probe state. This aligns with the behavior of non-Markovianity, which begins to saturate around this time. Once \(\mathcal{N}(t)\) reaches a steady state, no further information exchange occurs, and the probe attains the highest QFI. A similar observation has been reported in Ref.~\cite{PhysRevA.99.062114}, though without explicitly attributing it to the role of non-Markovianity. Moreover, at this stage, the QFI of the probe state significantly exceeds that of a single qubit in the steady state. Clearly, the probe system's ability to accumulate information over extended periods enables it to achieve substantially higher thermometric sensitivity.

\subsection{Estimation of ancilla qubit frequency}
\begin{figure}[t!]
    \centering
    \subfloat[]{\includegraphics[scale=0.34]{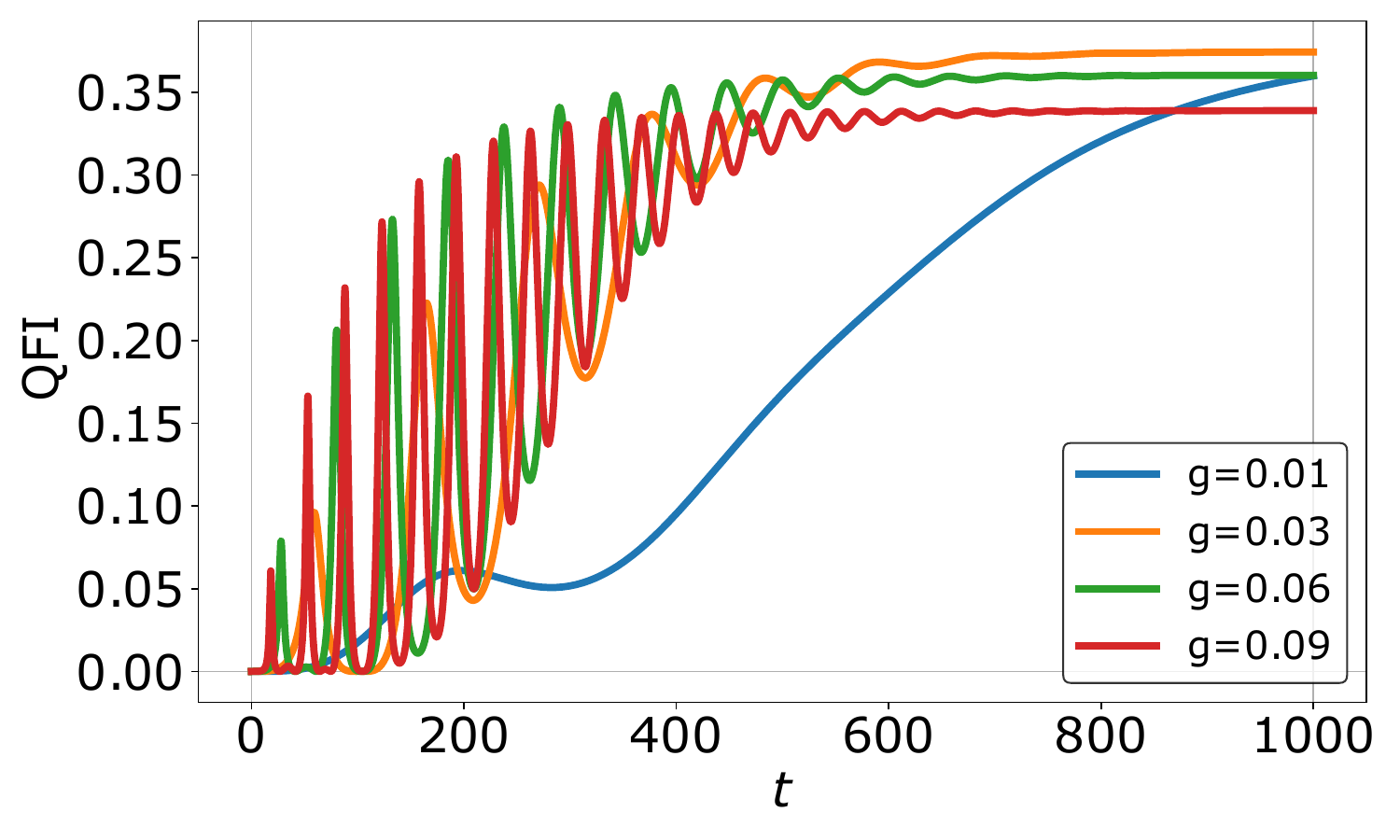}}\\
    \subfloat[]{\includegraphics[scale=0.35]{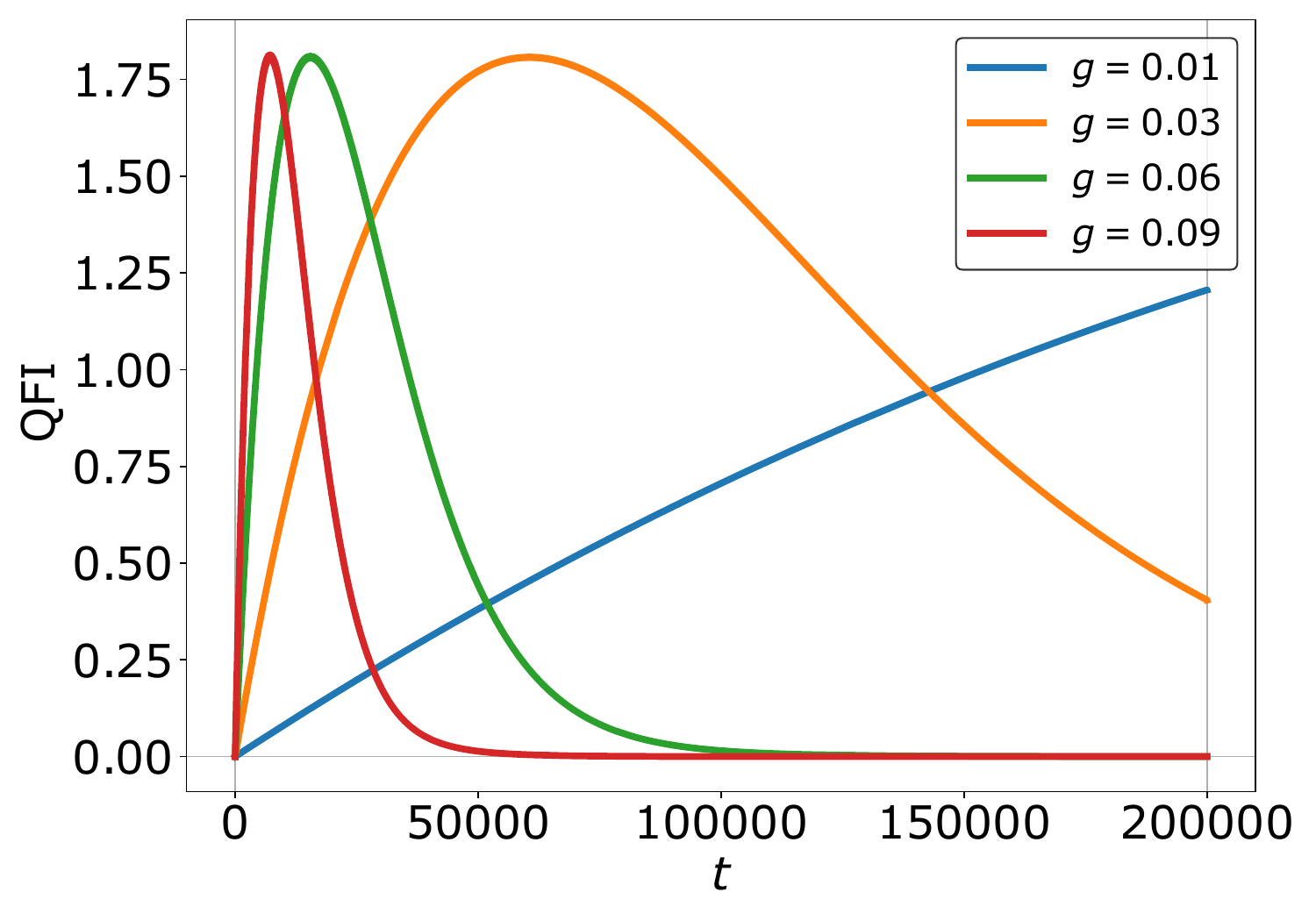}}
    \caption{QFI $\mathcal{F}_{\omega_A}$ as a function of time for estimation of frequency $\omega_A$ for (a) \( H^{XX}_\text{AP} \)  and (b) \( H^{ZX}_\text{AP} \) interaction. The parameter are set to $\omega_P=1$, $\omega_A=0.99$, $T=0.3$, and $\gamma=0.05$.}
    \label{fig:4}
\end{figure}

In what follows, we focus on estimating the frequency \(\omega_A\), assuming the temperature $T$ of the reservoir is known. This allows us to isolate the role of $\omega_A$ in the system dynamics and simplify the estimation problem. The frequency $\omega_A$ plays a critical role in structured reservoirs, as it corresponds to the central frequency of the Lorentzian spectral density that characterizes the reservoir. This central frequency governs the interaction channel between the probe and the bath, making its estimation essential for understanding and controlling the information exchange in the system. By employing different types of interaction Hamiltonians, we explore how $\omega_A$ can be accurately determined through measurements on the probe qubit.
In Fig.~\ref{fig:4}(a), we plot the QFI as a function of time for different values of coupling strength \( g \) under the \( H^{XX}_\text{AP} \) interaction. As observed in the previous section, the QFI increases very slowly towards its maximum value in the weak coupling regime (\( g=0.01 \)). However, as the coupling strength \( g \) increases, the QFI exhibits rapid oscillations and reaches a steady-state value more quickly. 

On the other hand, when we consider the $H^{XZ}_\text{AP}$ interaction (see Fig.~\ref{fig:4}(b)), the QFI is significantly higher and takes a much longer time to reach its maximum value in the weak coupling regime. However, as the coupling strength $g$ increases, the QFI reaches its peak more rapidly before eventually decaying to zero.  
Notably, for the $H^{ZX}_\text{AP}$ interaction, the QFI remains negligible for estimating the frequency $\omega_A$. These results indicate that the $H^{XZ}_\text{AP}$ interaction is the preferable choice for achieving higher precision in estimating the parameter $\omega_A$.
\begin{figure*}[t!]
    \centering
    \includegraphics[width=.9\textwidth]{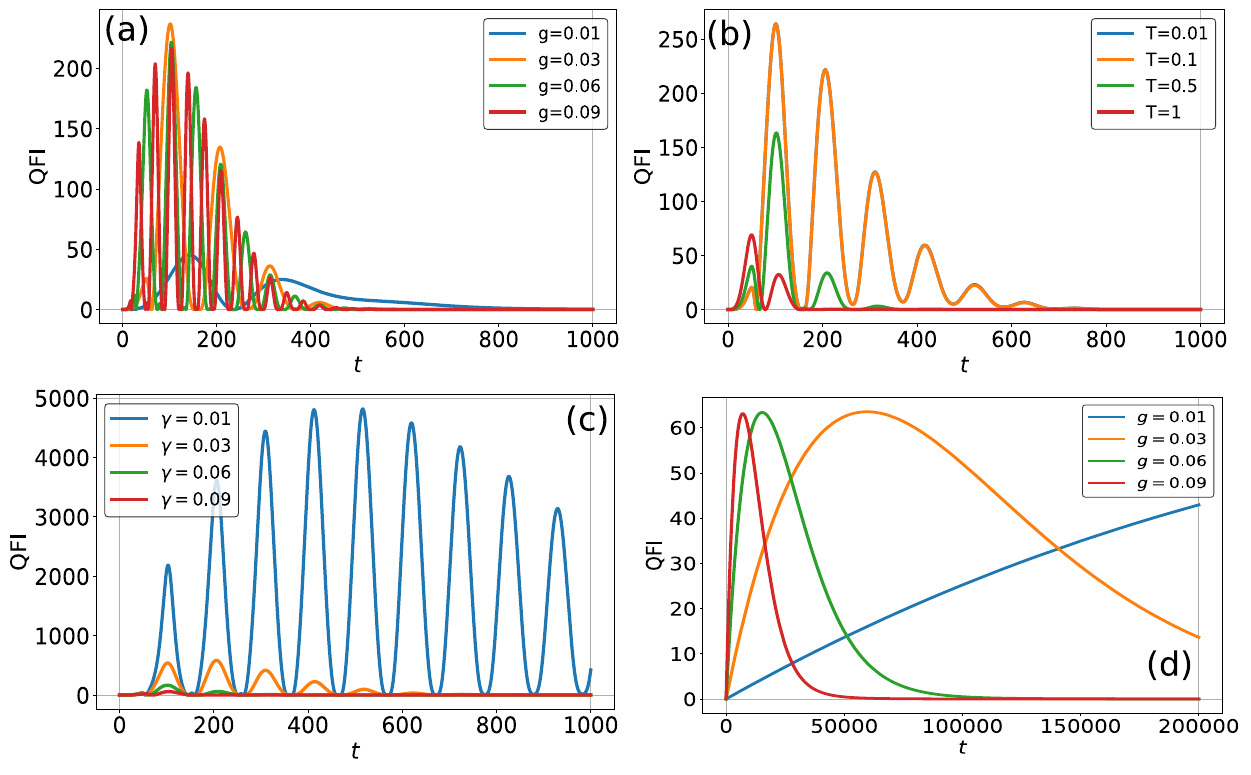}
    \caption{QFI $\mathcal{F}_{\gamma}$ as a function of time for estimation of ancilla qubit-reservoir coupling strength $\gamma$ for different values of (a) ancilla-probe coupling strength $g$, (b) reservoir temperature, and (c) ancilla qubit-reservoir coupling value $\gamma$. The interaction Hamiltonian between ancilla and probe is assumed to be \( H^{XX}_\text{AP} \). (d) The QFI $\mathcal{F}_{\gamma}$  as a function of $t$ for the interaction term \( H^{XZ}_\text{AP} \) for different values of coupling strength $g$. All the other parameters are set to $\omega_P=1$, $\omega_A=0.99$.}
    \label{fig5}
\end{figure*}
\subsection{Estimation of the ancilla qubit-reservoir coupling strength}
Finally, we present the results for the estimation of the ancilla qubit-reservoir coupling strength \(\gamma\). Figure.~\ref{fig5}(a) illustrates the QFI as a function of time for estimating \(\gamma\) for different values of the coupling strength \(g\) for \( H^{XX}_\text{AP} \) interaction. The QFI values for \(\gamma\) estimation are significantly larger compared to those for temperature (\(T\)) or frequency (\(\omega_A\)) estimation. For \( g=0.01 \), the QFI is initially small, reaches a maximum, and then decays to zero. However, for larger values of \(g\), the QFI exhibits rapid oscillations at the beginning of the evolution and decays much faster compared to the weak coupling case. Notably, the peak QFI value is highest for \( g=0.03 \), suggesting an optimal regime for estimating $\gamma$. These results underscore the critical role of $\gamma$ in mediating the interaction between the ancilla and the reservoir, as well as its influence on the non-Markovian dynamics of the system.

Based on this, we fix \( g=0.03 \) and analyze the QFI for different reservoir temperatures \(T\), as shown in Fig.~\ref{fig5}(b). Similar to the case of varying \( g \), the maximum QFI value remains unchanged across different values of \(T\) and eventually decays to zero over long time intervals. Furthermore, we examine the effect of varying \(\gamma\) on the QFI in Fig.~\ref{fig5}(c). It is observed that the QFI attains a very large value when the system-bath coupling strength is weak, such as \(\gamma=0.01\). In this regime, the QFI exhibits periodic oscillations and decays to zero very slowly. However, stronger values of \(\gamma\) have a detrimental effect on the QFI behavior.  

These results suggest that to achieve maximum sensitivity in estimating \(\gamma\), the system-reservoir coupling should be weak. Another important observation is that the QFI does not saturate, implying that information about \(\gamma\) can be extracted only in the transient regime. In the steady-state regime, the probe state becomes independent of the system-bath coupling information.  

The results for the \( H^{XZ}_\text{AP} \) interaction are presented in Fig.~\ref{fig5}(d) for different values of \( g \). Interestingly, the QFI for estimating \(\gamma\) exhibits lower sensitivity compared to its sensitivity for estimating other parameters such as \( T \) and \( \omega_A \). In the estimation of all parameters, the QFI for the \( H^{XZ}_\text{AP} \) interaction takes a significantly long time to reach its maximum value. However, this process is even slower for very weak coupling strengths, such as \( g=0.01 \) (solid blue curve in Fig.~\ref{fig5}(d)).  

Moreover, the \( H^{ZX}_\text{AP} \) interaction proves ineffective in estimating any other parameter of the structured reservoir. From these observations, we conclude that for the estimation of temperature \( T \) and frequency \( \omega_A \), the \( H^{XZ}_\text{AP} \) interaction is the most effective for extracting information. On the other hand, for estimating the coupling strength \(\gamma\), the \( H^{XX}_\text{AP} \) interaction outperforms the other types of interactions.

\subsection{Steady-state results for parameter estimation of the structured reservoir}
Since the results in the non-equilibrium regime show that parameter estimation accuracy depends on the nature of the probe-ancilla interaction, it is essential to investigate how these interactions perform in the steady-state regime. In particular, we will focus on the $ H^{(XX)}_{AP} $ and $ H^{(XZ)}_{AP} $ interactions, which have shown superior performance for the estimation of certain parameters in the non-equilibrium case. By analyzing the QFI in the steady state regime, we aim to determine whether these interactions maintain their advantage and how thermalization affects the extractable information.\par
To this end, we solve the Lindblad master equation (\ref{eq:ME}) analytically in the long-time limit ($t \to \infty$), by imposing the stationarity condition $\frac{d\rho}{dt} = 0$. This allows us to determine the steady-state density matrix $\rho_{\text{ss}}$ of the total system. From this solution, we compute the reduced state of the probe using Eq.~\eqref{Probe_state}, which serves as the basis for evaluating the steady-state QFI.
Under the $H^{(XZ)}_{AP}$ interaction, the probe’s reduced state in the steady-state limit becomes a maximally mixed state, given by:
\begin{eqnarray}
	\rho_P(\infty)= \frac{1}{2} \left(
\begin{array}{cc}
	1 & 0 \\
	0 & 1 \\
\end{array}
\right) 
\end{eqnarray}
In the steady-state regime, the probe loses all parameter information and becomes a maximally mixed state, fundamentally limiting the extractable information about the parameters of interest. As a result, the QFI for any parameter estimation vanishes, making the \( H^{(XZ)}_{AP} \) interaction ineffective for metrological purposes in equilibrium, unlike in the transient regime.
However, the scenario changes with the \( H^{(XX)}_{AP} \) interaction. Here, the steady-state density matrix of the probe retains a population imbalance, allowing for potential information retention. The probe state for this state is given by

\begin{equation}
\rho_P(\infty) = \begin{pmatrix}
	\frac{1}{2} + \Delta_p & 0 \\
	0 & \frac{1}{2} - \Delta_p
\end{pmatrix},
\end{equation}
where 
\begin{equation}
	\Delta_p = \frac{4 (\gamma_+ - \gamma_-) \omega_A \omega_P}{(\gamma_+ + \gamma_-) \left(8 g^2 + (\gamma_+ + \gamma_-)^2 + 4 (\omega_A^2 + \omega_P^2)\right)}.
\end{equation}
\begin{figure*}[t!]
	\includegraphics[width=1.01\textwidth]{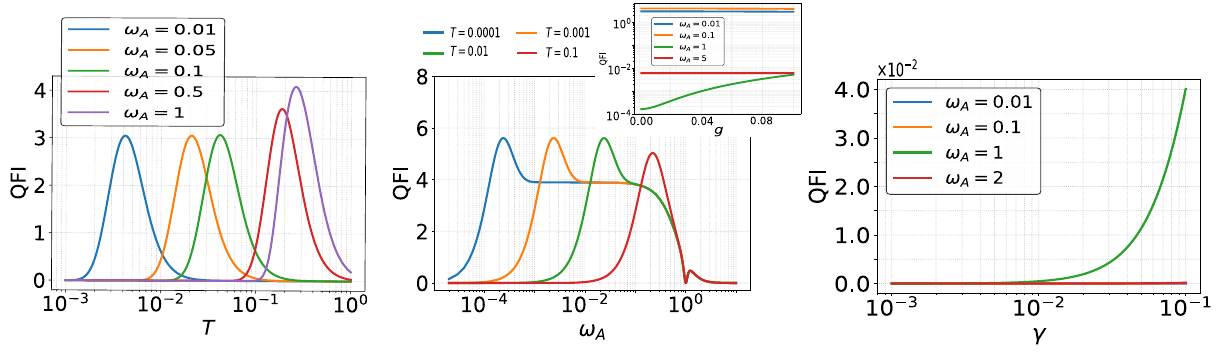}
	\put(-375,100){${(\textbf{a})}$}~
	\put(-320,100){${(\textbf{b})}$}~
	\put(-38,100){${(\textbf{c})}$}~
	\caption{Steady-state QFI for different parameters: (a) QFI with respect to $T$ for different values of $\omega_A$. (b) QFI with respect to $\omega_A$ for different values of $T$. The inset highlights the behavior of QFI as a function of $g$. (c) QFI with respect to $\gamma$ for various values of $\omega_A$. The parameters are set to  $\omega_P=1$, $T=0.01$, $g=0.08$, and $ \gamma=0.05 $.}
	\label{fig:Steady_state}
\end{figure*}
The term $\Delta_p$ represents the deviation from a maximally mixed state and reflects the asymmetry in the steady-state population distribution of the probe. This imbalance results from the difference in dissipation and excitation rates ($\gamma_+$ and $\gamma_-$), which are influenced by the bath temperature and the ancilla-bath interaction. As a result, $\Delta_p$ indicates that the steady state retains information about the interaction parameters of the system, making it relevant for parameter estimation in the equilibrium regime.\par
The explicit expressions for the QFI with respect to the temperature $T$, the ancilla frequency $\omega_A$, and the coupling strength $\gamma$ are respectively given by
\begin{eqnarray}
	\mathcal{F}_T(\infty) &=& \frac{4(A_T)^2}{1 - 4 \Delta_p^2},\label{EX_QFI_T}\\
	\mathcal{F}_{\omega_A}(\infty) &=& \frac{4(A_{\omega_A})^2}{1 - 4 \Delta_p^2},\label{EX_QFI_w}\\
	\mathcal{F}_{\gamma}(\infty) &=& \frac{4(A_{\gamma})^2}{1 - 4 \Delta_p^2},\label{EX_QFI_gamma}
\end{eqnarray}
where 
\begin{eqnarray}
	A_T&=&\frac{4 \omega _A^2 \omega _P \tanh ^2\left(\frac{\omega _A}{2 T}\right) \left(3 \gamma ^2-8 g^2+\xi -\Omega \right)}{T^2 \chi },\nonumber\\
	A_{\omega_A}&=&-\frac{\omega _P e^{-\frac{\omega _A}{T}} \left(e^{\frac{\omega _A}{T}}-1\right){}^2 \Lambda~ \text{sech}^2\left(\frac{\omega _A}{2 T}\right)}{2 T \chi },\nonumber\\
	A_{\gamma}&=&\frac{4 \gamma  \omega _A \omega _P \left(\sinh \left(\frac{2 \omega _A}{T}\right)-2 \sinh \left(\frac{\omega _A}{T}\right)\right)}{\chi }.
\end{eqnarray}
In the above expressions, certain terms are explicitly defined as
\begin{eqnarray}
	\Omega&=&4 \left(\omega _A^2+\omega _P^2\right),\nonumber\\
	\chi&=&\left(\left(\gamma ^2+8 g^2+\Omega \right) \cosh \left(\frac{\omega _A}{T}\right)+\gamma ^2-8 g^2-\Omega \right){}^2,\nonumber\\
	\xi&=& \left(3 \gamma ^2+8 g^2+\Omega \right) \cosh \left(\frac{\omega _A}{T}\right),
\end{eqnarray}
and
\begin{widetext}
	\begin{equation}
		\Lambda=2 \omega _A \left(3 \gamma ^2-8 g^2+\xi -\Omega \right)+4 T \sinh \left(\frac{\omega _A}{T}\right) \left(4 \omega _A^2+\gamma ^2-8 g^2-4 \omega _P^2\right).
	\end{equation}	
\end{widetext}

Figure \ref{fig:Steady_state}(a) shows the steady-state QFI (Eq. \eqref{EX_QFI_T}) as a function of temperature T for different values of the ancilla qubit frequency $\omega_A$. These results show a characteristic peak indicating an optimum temperature at which the information about T is maximized. For small values of $\omega_A$, this peak appears at ultra-low temperatures, while for larger $\omega_A$ it shifts to low temperatures. This behavior is consistent with the analytical expression of the QFI, where temperature information is encoded in the population asymmetry $\Delta_p$, which depends on the dissipation rates influenced by the bath temperature. More specifically, $\Delta_p$ disappears at higher temperatures where $\gamma_+ \approx \gamma_-$, meaning that the population imbalance disappears and the probe state approaches a maximally mixed state.\par

The middle panel (Fig. \ref{fig:Steady_state}(b)) shows QFI (Eq. \eqref{EX_QFI_w}) with respect to $\omega_A$ for estimation of $\omega_A$. The plot shows the effect of temperature on the range of $\omega_A$ values where the estimation is most accurate. At lower temperatures ($T = 0.0001$, blue curve), the QFI remains significant over a wider range of $\omega_A$, allowing a more flexible estimation range. However, as the temperature increases ($T = 0.1$, red curve), the QFI becomes more localized around certain $\omega_A$ values, effectively reducing the range where a high-precision estimate is possible. This suggests that thermal fluctuations play a crucial role in limiting the frequency range suitable for estimation, making low-temperature regimes more favorable for robust metrological performance.\par
The inset plot further illustrates the role of probe-ancilla coupling $g$ in the estimation of $\omega_A$. In the off-resonance regime ($\omega_A \neq 1$), the QFI remains nearly stationary with respect to $g$, indicating that variations in coupling strength have only a minor effect on estimation accuracy. However, as the ancilla frequency gets closer to the probe frequency ($\omega_A = 1$), the QFI becomes highly sensitive to $g$.\par

Finally, Fig. \ref{fig:Steady_state}(c) shows that the QFI (Eq. \eqref{EX_QFI_gamma}) for estimating the coupling strength \( \gamma \) between the ancilla and the bath exhibits different behavior in the resonance and off-resonance regimes.
In the off-resonance case, the QFI is negligible and remains almost stationary with respect to $\gamma$, reflecting a very low sensitivity. However, in the resonance regime $(\omega_A = 1) $, the QFI becomes sensitive to $\gamma$, increases significantly with increasing coupling strength. Despite this sensitivity, the magnitude of the QFI in the resonance regime is much smaller than in the non-equilibrium regime, emphasizing the limitations of steady-state systems in achieving high precision.
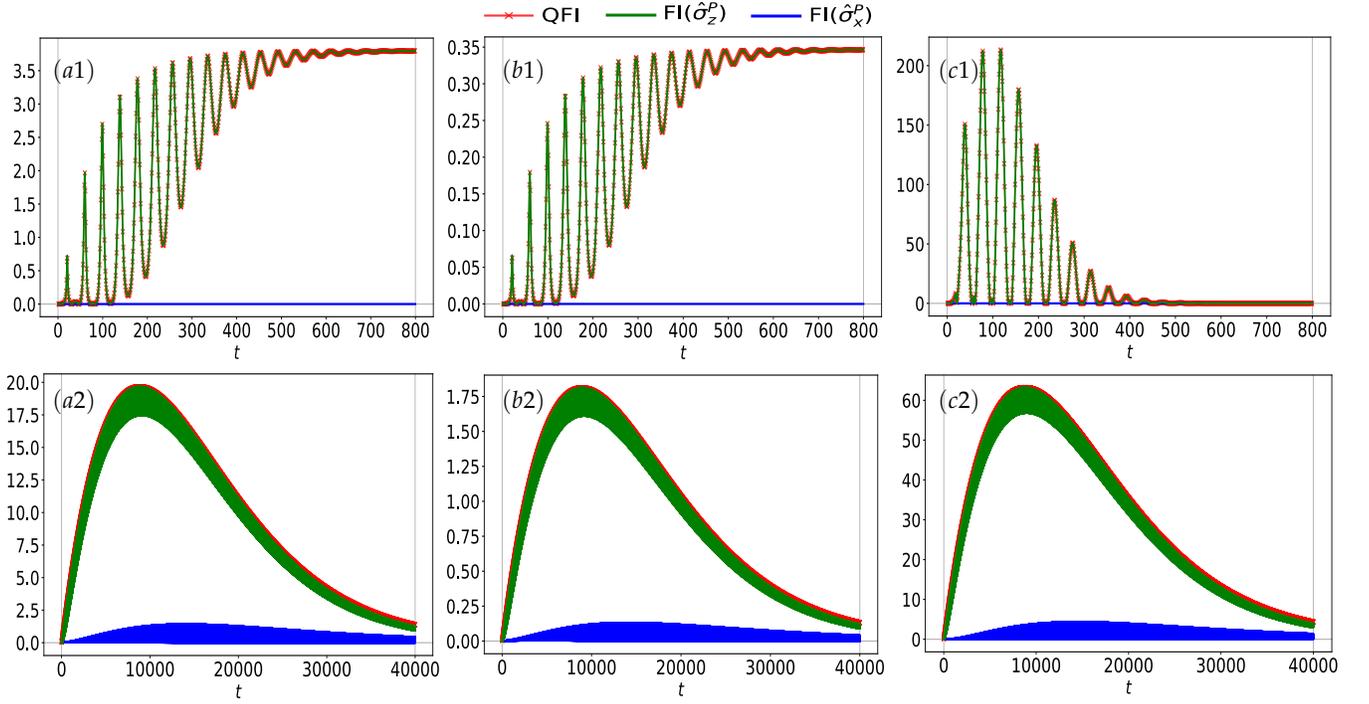
\begin{figure*}[t!]
	\includegraphics[width=1.0\textwidth]{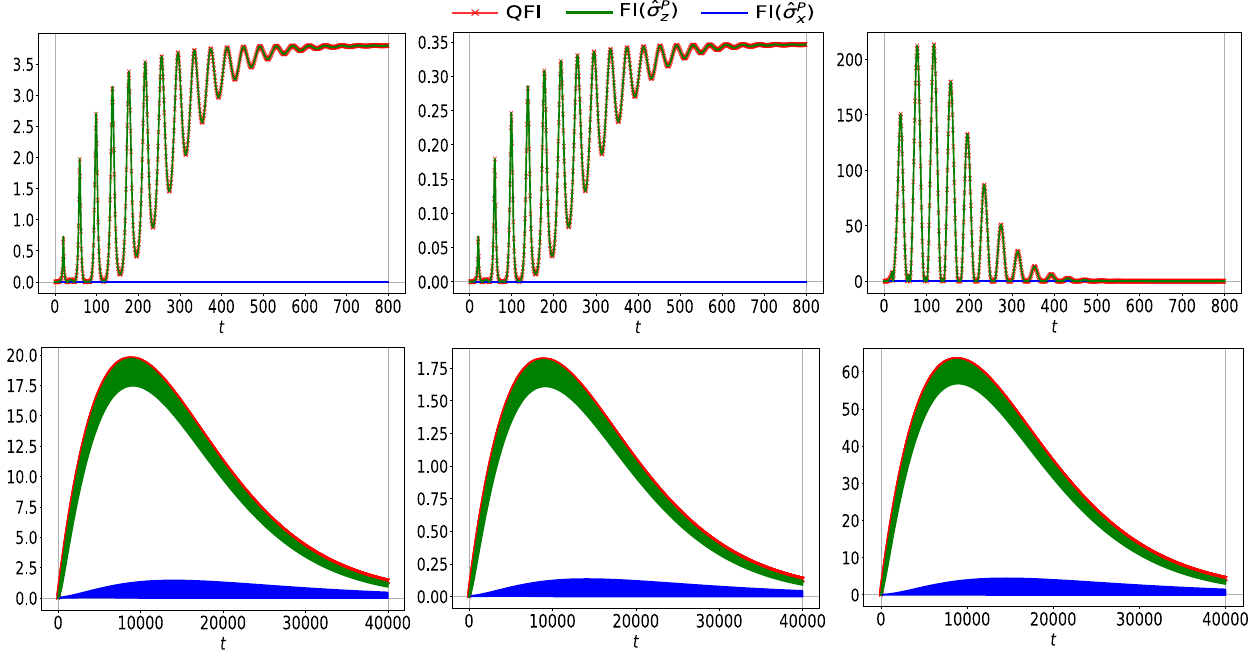}
	\put(-490,240){${(a1)}$}~
	\put(-490,115){${(a2})$}~
	\put(-320,240){${(b1)}$}~
	\put(-320,115){${(b2)}$}~
	\put(-155,240){${(c1)}$}~
	\put(-155,115){${(c2})$}~	
	\caption{Comparison of QFI and FI for measurements of $\hat{\sigma}_z^P$ and $\hat{\sigma}_x^P$ in $H^{(XX)}_{AP}$ (top) and $H^{(XZ)}_{AP}$ (bottom) interactions. The columns represent parameter estimation for: (a1, a2) temperature ($T$), (b1, b2) ancilla frequency $\omega_A$, and (c1, c2) coupling strength $\gamma$. Parameters are fixed at $\omega_A = 0.99$, $\omega_P = 1$, $T = 0.3$, $g = 0.08$, and $\gamma = 0.05$.}
	\label{fig:8}
\end{figure*}
\section{conclusion}
\label{conclusion}
In this paper, we present a scheme for estimating key parameters of a structured reservoir using a two-level quantum probe. The structured reservoir is made up of an ancilla qubit coupled to a thermal bath at temperature $T$, where the ancilla and the bath together form a structured environment for the probe. A structured reservoir is characterized by a spectral density function typically modeled as a sum of Lorentzians. In our scheme, the ancilla frequency $\omega_A$ corresponds to the central frequency of the Lorentzian, while the coupling strength $\gamma$ and temperature $T$ determine the structured reservoir's influence on the probe.
The probe qubit interacts with the bath indirectly through the ancilla, enabling the use of the probe-ancilla interaction to estimate the parameters of the structured reservoir: the bath temperature \( T \), the ancilla-bath coupling strength \( \gamma \), and the ancilla’s transition frequency \( \omega_A \).
To analyze the dynamics of the system, we use the Lindblad master equation, which describes the dissipative evolution of the system-bath interaction, while allowing the tracing of the ancilla to focus on the reduced state of the probe. We explore different interaction Hamiltonians between the probe and the ancilla (\( H^{XX}_{AP} \), \( H^{XZ}_{AP} \), \( H^{XX+ZX}_{AP} \), and \( H^{ZX}_{AP} \)) and demonstrate how the choice of interaction significantly affects parameter estimation precision.
Non-Markovianity is crucial in the transient regime, allowing sustained information backflow that improves the QFI for temperature and frequency estimation.  
Notably, the $ H^{XZ}_{AP} $ interaction exhibited the highest non-Markovianity and generated significant coherence in the probe, which persisted over extended periods, making it the most effective for temperature and frequency estimation.  

The other two interactions, $ H^{XX}_{AP} $ and $ H^{XX+ZX}_{AP} $, provide nearly identical information for temperature estimation, with maximum sensitivity reached as the probe approaches its steady state. Non-Markovian effects in both interactions manifest oscillations and accelerated thermalization of the probe, leading to faster attainment of maximum QFI.
In contrast, the $ H^{ZX}_{AP} $ interaction proved ineffective for parameter estimation due to its antisymmetric nature, which caused dissipation and prevented efficient information encoding. The \( H^{XX}_{AP} \) interaction, though less effective for temperature and frequency estimation, outperformed in estimating the coupling strength \( \gamma \), especially in the weak coupling regime.

Moreover, a comparison between transient and steady-state regimes revealed distinct advantages depending on the parameter of interest and interaction type. In the transient regime, non-Markovian effects allowed for high-precision estimation, with the $ H^{XZ}_{AP} $ interaction achieving the highest QFI values for temperature and frequency estimation. However, in the steady-state regime, the $ H^{XZ}_{AP} $ interaction led to a maximally mixed probe state, rendering its ineffectiveness for quantum metrology. On the other hand, the $ H^{XX}_{AP} $ interaction retained a population imbalance in the steady state, enabling parameter estimation even at equilibrium, though with reduced precision compared to the transient regime.  

Our results emphasize the importance of selecting appropriate interactions and operating regimes for optimal parameter estimation. The interplay between non-Markovianity, coherence generation, and interaction type provides valuable insights into quantum metrology and the characterization of structured reservoirs, thereby enabling improved precision in quantum sensing applications.

Finally, although our work is theoretical in nature, the proposed system is suitable for practical implementation in several quantum platforms~\cite{scigliuzzo2020primary,Viisanen_2015}, particularly in circuit quantum electrodynamics (cQED) architectures. Such a setup can be experimentally implemented using nano-devices~\cite{PhysRevA.97.032133} operating in the quantum regime, such as flux qubits, Cooper pair boxes, or transmon qubits. In a typical realization, one qubit (the probe) is coherently coupled to a second qubit (the ancilla), which in turn is coupled to a dissipative element—such as a resistor or a lossy resonator—acting as a thermal reservoir. This configuration effectively creates a structured reservoir and enables precise engineering of the system-environment interaction.

\section*{ACKNOWLEDGMENTS}
A.U. and \"O.E.M. acknowledge support from the Scientific and Technological Research Council (T\"UBİTAK) of T\"urkiye under Project Grant No. 123F150. Y.A. acknowledges the support of the French Government through the \textit{Programme de Mobilités Doctorales} (No. 166019X) during his research stay in France. Y.A. and A.U. contributed equally to this work.
\appendix
\section{Optimal measurements}
\label{Optimal_measurements}

In this appendix, we explore the optimal measurement strategy for extracting information about key parameters of the structured reservoir, which consists of an ancilla qubit coupled to a thermal bath. Specifically, we focus on estimating the bath temperature (\( T \)), the coupling strength between the ancilla and the bath (\( \gamma \)), and the ancilla’s frequency (\( \omega_A \)) using a probe qubit.

To determine the most effective measurement, we compare the FI obtained from specific projective measurements with the QFI, which represents the fundamental upper bound on achievable precision. In particular, we will analyze the measurements in the eigenbases of $ \hat{\sigma}_x $ and $ \hat{\sigma}_z $ to assess how well they capture the relevant information. This comparison allows us to identify the most informative measurement basis and determine whether a simple projective measurement is sufficient to reach the optimal sensitivity predicted by QFI.

When a measurement is performed on the system, it yields an outcome $\epsilon$ that follows a probability distribution $p(\epsilon | \theta)$. This distribution depends on an unknown parameter $\theta$, which we aim to estimate. The FI quantifies how much information about $\theta$ can be extracted from the measurement and is given by~\cite{paris2009}:
\begin{equation}
	\text{FI} = \int d\epsilon \, p(\epsilon|\theta) \left( \frac{\partial \ln p(\epsilon|\theta)}{\partial \theta} \right)^{2}.
\end{equation}
For projective measurements on a two-level system, this expression simplifies to:
\begin{equation}
	\text{FI}(\hat{X}) = \frac{1}{\langle \Delta \hat{X}^{2} \rangle} \left( \frac{\partial \langle \hat{X} \rangle}{\partial \theta} \right)^{2},
\end{equation}
where $\langle \hat{X} \rangle$ is the expectation value of the observable $\hat{X}$, and $\langle \Delta \hat{X}^{2} \rangle$ is its variance.

Figure~\ref{fig:8} shows a comparison of the QFI and FI associated with projective measurements on the probe qubit in the $ H_{AP}^{(XX)}$ and $ H_{AP}^{(XZ)}$ interaction settings. The top row (a1, b1, c1) corresponds to measurements under the $ H_{AP}^{(XX)}$ interaction, while the bottom row (a2, b2, c2) corresponds to the $ H_{AP}^{(XZ)}$ interaction. Each column represents a different estimation task: temperature ($T $, left), ancilla frequency ($\omega_A $, middle) and coupling strength ($\gamma $, right).

For $ H^{(XX)}_{AP} $ interactions (top row), FI from the measurement of $ \hat{\sigma}_z^P $ (green) precisely follows QFI (red), indicating that population measurements are sufficient to saturate the quantum limit, while FI from the measurement of $ \hat{\sigma}_x^P $ (blue) remains zero, showing that coherences do not contribute to parameter estimation. This suggests that the information is fully encoded in the populations. In contrast, for $H^{(XZ)}_{AP} $ interactions (bottom row), FI from $\hat{\sigma}_x^P $ is non-zero, meaning that coherences now play a role in encoding information, but FI from $\hat{\sigma}_z^P $ still dominates and closely follows QFI. The oscillatory behavior in FI suggests a periodic flow of information back and forth between the system and the environment, improving the accuracy of estimates at certain times. This highlights the fundamental difference between the two interaction types, where $ H^{(XX)}_{AP} $ interactions restrict useful information to populations, while $ H^{(XZ)}_{AP} $ interactions allow coherence to contribute, although sub-dominantly.
\section{Role of coherence} \label{coh}   
\begin{figure}[t!]
        \centering
        \includegraphics[scale=0.33]{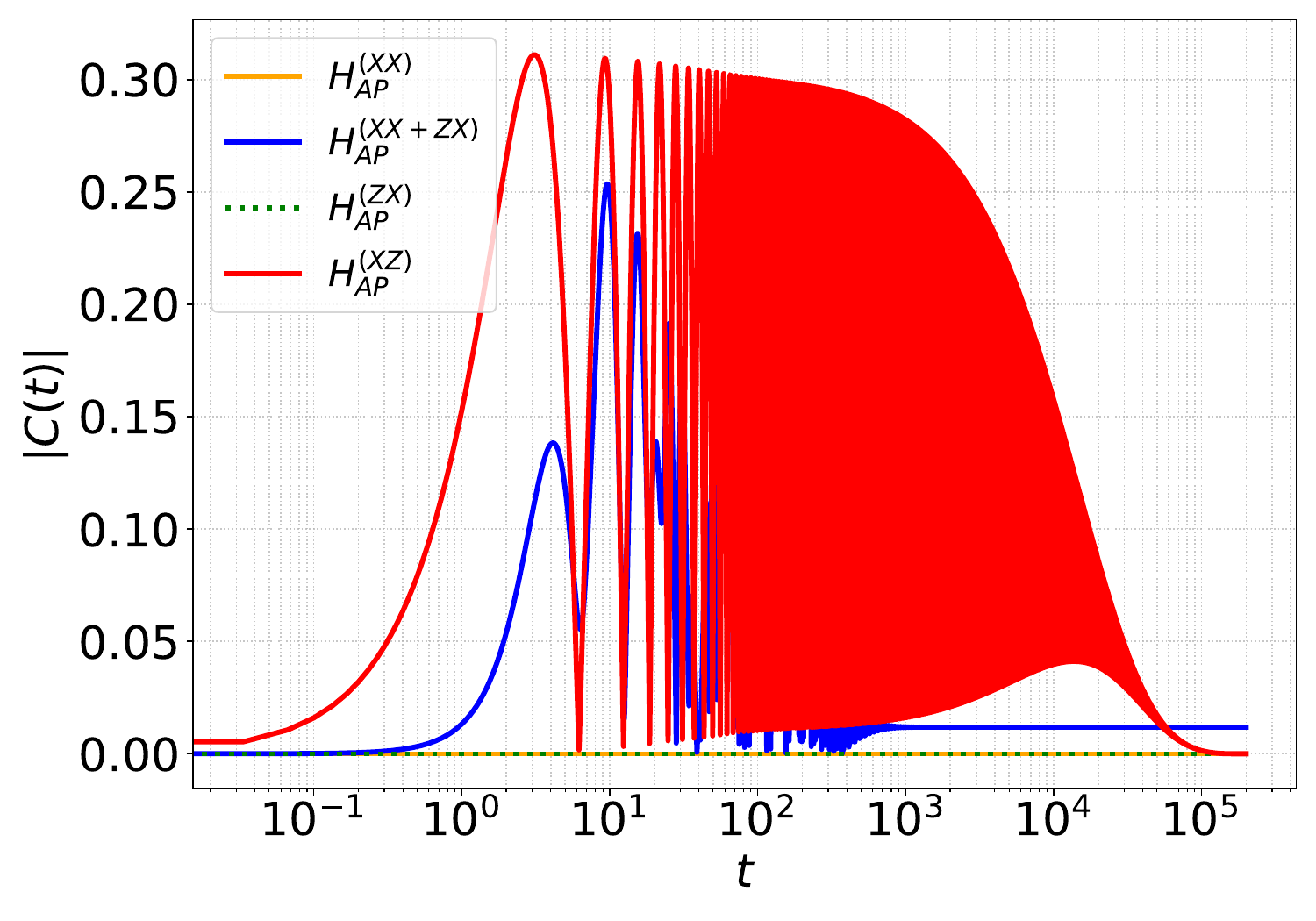}
        \caption{The absolute value of coherence $|C(t)|$ of the excited state of population of the probe qubit as a function of time $t$ for different models. The parameters are set to $\omega_P=1$. $\omega_A=0.99$, $T=0.3$, $\gamma=0.05$, and $g=0.08$.}
        \label{fig:coh}
    \end{figure}

Quantum coherence plays a central role in quantum mechanics and quantum computing, as it quantifies the ability of a qubit to maintain its quantum state over time. In this work, we consider the \(l_1\)-norm measure of coherence, defined as
   \begin{equation}
   C_{l_1}(\rho) = \sum_{i \neq j} |\rho_{ij}|
   \end{equation}
where \(\rho_{ij}\) are the off-diagonal elements of the density matrix \(\rho\). This measure captures the magnitude of the off-diagonal elements of the density matrix of the probe, which represent quantum coherence. We investigate the quantum coherence of the probe state as a function of time for different types of interactions. In Fig.~\ref{fig:coh}, we plot the \( l_1 \)-norm of coherence over time. 

First, we observe that the coherence for the individual interactions \( H^{XX}_\text{AP} \) and \( H^{ZX}_\text{AP} \) remains zero at all times, as shown by the orange and green dotted curves. However, when these two interactions are combined, the coherence becomes non-zero. This phenomenon arises because the combined \( H^{XX+ZX}_\text{AP} \) interaction introduces additional terms in the Hamiltonian structure, which can generate coherence. 
On the other hand, for the \( H^{XZ}_\text{AP} \) interaction, the maximum coherence value is slightly higher than that of \( H^{XX+ZX}_\text{AP} \) in Fig.~\ref{fig:coh}. Notably, the coherence persists for a significantly longer time in this case. This underscores the crucial role of \( H^{XZ}_\text{AP} \) interactions in sustaining coherence over extended periods. Such interactions are particularly valuable in quantum computation, where maintaining coherence is essential for minimizing decoherence and ensuring reliable quantum operations.
\bibliography{Mods.bib}
\end{document}